\documentclass{article}
\usepackage{spconf,amsmath,amssymb,graphicx}
\usepackage[hidelinks]{hyperref}
\usepackage{float}
\usepackage{placeins}
\usepackage{fancyhdr}

\title{TrustFlow-ECG: Reliability-Aware ECG Restoration with Calibrated Risk Estimation}
\name{Mohamed Kamel, Tamer Nadeem}
\address{Department of Computer Science, Virginia Commonwealth University, Richmond, VA, USA\\
\{kamelms,tnadeem\}@vcu.edu}
\begin{document}
%
\maketitle
\thispagestyle{fancy}
\begin{abstract}
Electrocardiogram (ECG) restoration models can reduce noise and motion artifacts, but they rarely estimate whether the restored waveform is reliable for downstream analysis. We propose TrustFlow-ECG, a reliability-aware restoration framework that combines a one-step conditional rectified-flow restorer with a lightweight window-level risk predictor. The predictor uses the noisy ECG, restored ECG, and correction residual, then applies validation-set isotonic calibration for selective ECG analysis. We evaluate on MIT-BIH ECG windows corrupted with NSTDB artifacts, powerline interference, and mixed noise across multiple SNRs, with an external 9{,}675-window PTB-XL stress check without retraining. TrustFlow achieves 0.862 Spearman correlation with true risk and 0.911 AUROC for failure detection, while calibration reduces ECE from 0.0385 to 0.0098. At 80\% coverage, selective rejection reduces RMSE from 0.1848 to 0.1345, improves R-peak F1 from 0.8778 to 0.9175, and reduces HR MAE from 6.26 to 4.88 bpm.
\end{abstract}


\section{Introduction}
\label{sec:intro}

Electrocardiogram (ECG) recordings are widely used for cardiac rhythm analysis, heart-rate estimation, and long-term physiological monitoring. In ambulatory or wearable settings, baseline wander, muscle artifact, electrode motion, and powerline interference can distort ECG morphology and beat timing. Public resources such as PhysioNet and the MIT-BIH Arrhythmia Database support reproducible ECG algorithm evaluation, while classical QRS detection remains a common basis for beat-level assessment~\cite{Goldberger2000PhysioNet,932724,4122029}.

ECG denoising and restoration methods include signal-decomposition approaches, convolutional and transformer-based models, and diffusion or score-based generative models~\cite{10889014,10018543,10891895}. Generative and flow-based methods are also attractive for biosignal and audio restoration because they model complex signal distributions while supporting efficient one- or few-step sampling~\cite{Lipman2022FlowMF,Liu2022FlowSA,Tong2023ConditionalFM}. However, these methods mainly optimize average waveform quality using metrics such as RMSE, correlation, SNR improvement, or beat-detection accuracy.

Average restoration accuracy is not enough for reliability-critical ECG analysis. A restored window may look plausible while still containing morphology errors or shifted R-peaks that affect downstream measurements. Existing ECG quality methods estimate whether the observed input is suitable for analysis~\cite{10491116,HUANG2024107928}, while recent ECG systems estimate uncertainty for diagnostic or classification outputs~\cite{10.1093/ehjdh/ztab045,DEMOLDER202471}. In contrast, we estimate the reliability of the \emph{restored ECG output itself}: can the system predict whether a restored window should be trusted?

This problem is related to calibration and selective prediction, where confidence is aligned with observed error and used to reject low-reliability examples~\cite{3305381.3305518,3295222.3295241}. Recent work also emphasizes explicit coverage and risk evaluation for selective systems~\cite{3737916.3737992}, but these ideas are less explored for ECG restoration, where failures involve both waveform error and beat-timing degradation.

We propose \textit{TrustFlow-ECG}, a reliability-aware ECG restoration framework with calibrated risk estimation. A conditional rectified-flow backbone restores noisy ECG windows in one step, while a lightweight predictor estimates window-level risk from the noisy ECG, restored ECG, and correction residual. Clean ECG is used only during training and validation to define and calibrate risk.

\begin{figure*}[t]
    \centering
    \includegraphics[width=0.97\textwidth]{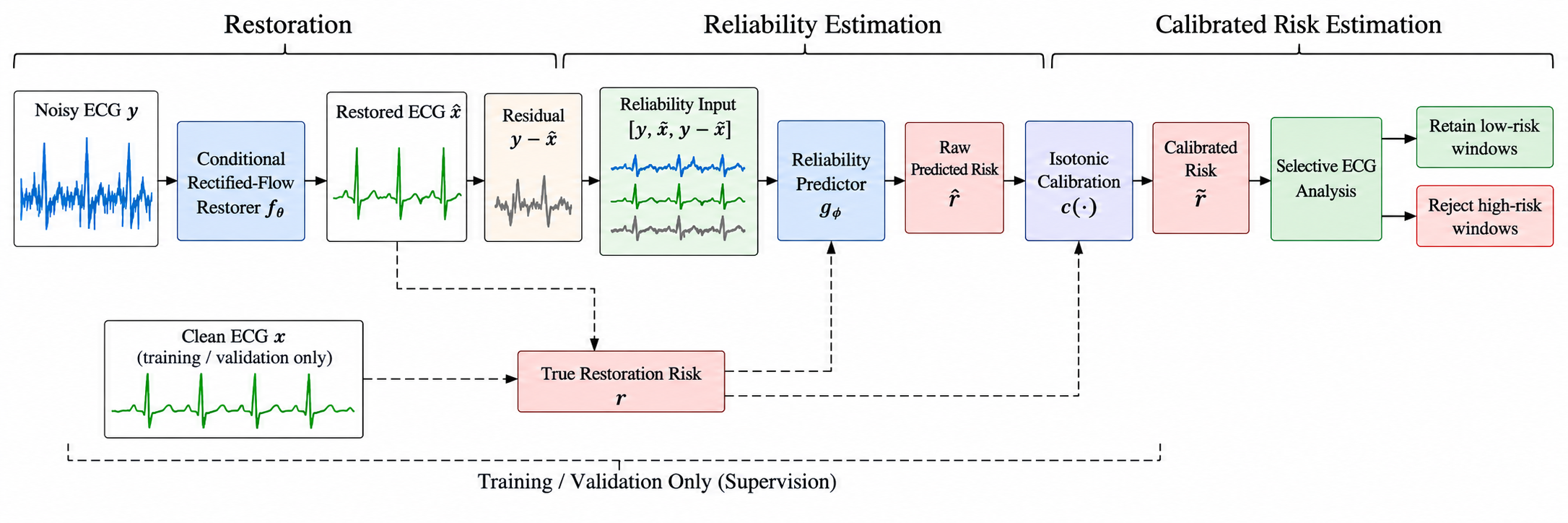}
    \caption{Overview of TrustFlow-ECG. A noisy ECG window is restored by a conditional rectified-flow model. The noisy ECG, restored ECG, and correction residual are used to predict restoration risk. Clean ECG is used only during training and validation to define true risk and fit calibration; at inference, calibrated risk guides selective ECG analysis.}
    \label{fig:framework}
\end{figure*}

The main contributions of this work are:
\begin{itemize}
    \item We formulate ECG restoration as a reliability-aware problem that restores the waveform and estimates when the restored window is unreliable.
    \item We introduce a window-level risk predictor using the noisy ECG, restored ECG, and correction residual without requiring clean ECG at inference.
    \item We calibrate predicted risk using validation data and evaluate ranking, failure detection, calibration, selective ECG analysis, and external PTB-XL transfer.
\end{itemize}

\section{Reliability-Aware ECG Restoration}
\label{sec:method}

Fig.~\ref{fig:framework} shows the TrustFlow-ECG framework. A noisy ECG window is first restored using a conditional rectified-flow backbone. A reliability predictor then estimates window-level restoration risk from the noisy ECG, restored ECG, and correction residual. Clean ECG is used only during training and validation to define true risk and fit calibration; at inference, the model uses only the noisy input and restored output.

\subsection{Problem formulation}
\label{subsec:problem}

Let $x \in \mathbb{R}^{T}$ denote a clean ECG window and $y \in \mathbb{R}^{T}$ its corrupted observation:
\begin{equation}
    y = x + \alpha n ,
    \label{eq:noisy_ecg}
\end{equation}
where $n$ is an artifact signal and $\alpha$ sets the target input SNR. Given $y$, a restoration model estimates $\hat{x}=f_{\theta}(y)$. Standard evaluation measures the discrepancy between $\hat{x}$ and $x$; here we also estimate whether the restored window is reliable. The reliability score is produced from inference-time signals,
\begin{equation}
    s = g_{\phi}(y,\hat{x},y-\hat{x}),
    \label{eq:risk_predictor_mapping}
\end{equation}
without observing clean ECG at inference. The score is mapped back to the original risk scale to obtain the predicted risk $\hat{r}$.

\subsection{Conditional flow restoration}
\label{subsec:flow}

We use a conditional rectified-flow backbone for ECG restoration. Rectified flow and flow matching learn a velocity field between two endpoints~\cite{Lipman2022FlowMF,Liu2022FlowSA,Tong2023ConditionalFM}. Here, the noisy ECG is the initial endpoint and the clean ECG is the target endpoint. For $t \sim \mathcal{U}(0,1)$,
\begin{equation}
    x_t = (1-t)y + tx, \qquad v^{\star}=x-y .
    \label{eq:flow_path}
\end{equation}
The network learns $v_{\theta}(x_t,t,y)$ to predict $v^{\star}$. At inference, restoration is performed with one endpoint update,
\begin{equation}
    \hat{x}=y+v_{\theta}(y,0,y).
    \label{eq:onestep_restoration}
\end{equation}
This keeps inference efficient while aligning restoration with a clean-endpoint transport from corrupted to clean ECG.

\subsection{Restoration-Risk Target}
\label{subsec:risk_target}

During training and evaluation, clean ECG is used to define a scalar restoration-risk target combining waveform error, local morphology error, R-peak preservation, and heart-rate error:
\begin{equation}
\begin{split}
    r =
    &\ \mathrm{RMSE}(\hat{x},x)
    + \lambda_d \frac{1}{T-1}\|\Delta \hat{x}-\Delta x\|_1  \\
    &+ \lambda_p(1-F^{R}_{1})
    + \lambda_h \min(e_{\mathrm{HR}}, h_{\max}),
\end{split}
\label{eq:composite_risk}
\end{equation}
where $\Delta$ is the first-order temporal difference operator, $F^{R}_{1}$ is the R-peak F1 score, and $e_{\mathrm{HR}}$ is the absolute heart-rate error in bpm. We use $\lambda_d=0.5$, $\lambda_p=0.1$, $\lambda_h=0.01$, and $h_{\max}=30$. This risk target is used only for supervision and evaluation, not at inference.

\subsection{Risk Prediction}
\label{subsec:reliability_prediction}

The reliability module predicts risk from signals available at inference:
\begin{equation}
    z = [y,\hat{x},y-\hat{x}] \in \mathbb{R}^{3 \times T}.
    \label{eq:reliability_input}
\end{equation}
The correction residual $y-\hat{x}$ helps expose how much and where the restoration model changed the input. The predictor $g_{\phi}$ is a lightweight one-dimensional CNN with strided convolutional blocks, global pooling, and a regression head. It is trained with a Huber loss on a standardized log-risk target,
\begin{equation}
\begin{aligned}
r^{\log} &=
\frac{\log(1+r)-\mu_{\log r}}{\sigma_{\log r}},\\
\mathcal{L}_{\mathrm{risk}}
&=
\mathrm{SmoothL1}\!\left(g_{\phi}(z),r^{\log}\right).
\end{aligned}
\label{eq:risk_loss}
\end{equation}
where $\mu_{\log r}$ and $\sigma_{\log r}$ are computed from the training split.

\subsection{Risk calibration and selective analysis}
\label{subsec:calibration_selective}

Raw predicted risk is useful for ranking, but it may not be calibrated to the observed risk scale. We therefore fit an isotonic calibration map $c(\cdot)$ on the validation set~\cite{3305381.3305518},
\begin{equation}
    \tilde{r}=c(\hat{r}).
    \label{eq:calibrated_risk}
\end{equation}
Calibration is evaluated using risk expected calibration error,
\begin{equation}
    \mathrm{ECE}_{r}
    =
    \sum_{b=1}^{B}
    \frac{|B_b|}{N}
    \left|
    \frac{1}{|B_b|}\sum_{i \in B_b} \tilde{r}_i
    -
    \frac{1}{|B_b|}\sum_{i \in B_b} r_i
    \right|,
    \label{eq:risk_ece}
\end{equation}
where $B_b$ denotes a bin of examples sorted by predicted risk.

For selective ECG analysis, we retain the $qN$ windows with the lowest calibrated risk and reject the rest:
\begin{equation}
    \mathcal{S}_{q}
    =
    \operatorname*{arg\,min}_{\mathcal{S}:|\mathcal{S}|=qN}
    \sum_{i\in\mathcal{S}} \tilde{r}_i .
    \label{eq:selective_set}
\end{equation}
Restoration metrics are recomputed only on $\mathcal{S}_{q}$, following the general idea of selective prediction~\cite{3295222.3295241}.

\section{Experimental Setup}
\label{sec:experiments}

\subsection{Dataset and noise protocol}
\label{subsec:dataset}

Our main evaluation uses clean ECG windows from the MIT-BIH Arrhythmia Database and artifact recordings from the MIT-BIH Noise Stress Test Database, both distributed through PhysioNet~\cite{Goldberger2000PhysioNet,932724}. Each example is a single-channel 10-second ECG window sampled at 125 Hz ($T=1250$). Corrupted inputs are generated by adding scaled artifact signals at SNRs of 0, 5, 10, 15, and 20 dB. This paired setup is needed because true restoration risk requires a clean reference ECG, which is usually unavailable in naturally noisy ambulatory recordings.

We evaluate baseline wander, muscle artifact, electrode motion, synthetic powerline interference, and mixed noise. We use record-level train/validation/test splits, with model selection and calibration performed only on validation records. Final main results use a fixed 4000-window MIT-BIH/NSTDB test manifest from eight held-out ECG records, five artifact types, five SNR levels, and 20 windows per record--noise--SNR condition.

For external transfer, we construct a PTB-XL stress set from single-lead 10-second ECG records resampled to 125 Hz and corrupted using the same artifact protocol~\cite{Wagner2020PTBXL}. This set contains 9{,}675 windows from 387 records. No PTB-XL examples are used for training, model selection, calibration, or threshold selection.

\subsection{Models and Evaluation Protocol}
\label{subsec:models_evaluation}

TrustFlow-ECG uses the conditional rectified-flow restorer from Eq.~\eqref{eq:onestep_restoration}. For restoration comparison, we include a one-dimensional U-Net and a transformer denoiser. These models provide restoration context, while the main contribution is the reliability layer built on top of the restored output.

For reliability evaluation, we compare TrustFlow risk against correction energy, oracle SNR ranking, and a handcrafted SQI-style regressor. Correction energy uses the residual magnitude and variation. Oracle SNR uses the true corruption level and is included only as an analysis reference. The SQI regressor uses 96 inference-time features from the noisy ECG, restored ECG, and correction residual, giving it the same inference-time signals as TrustFlow but without learned temporal feature extraction. We compare both raw and isotonic-calibrated neural risk scores.

The restoration model is trained on noisy--clean ECG pairs and then frozen. The reliability predictor is trained on restored outputs using the composite risk target in Eq.~\eqref{eq:composite_risk} and the input in Eq.~\eqref{eq:reliability_input}. The risk target is log-transformed, standardized using training-set statistics, and optimized with a Huber regression loss. Isotonic regression is fitted only on validation data; no test examples are used for model selection, calibration, or threshold selection. For PTB-XL, we reuse the MIT-BIH/NSTDB-trained models, validation calibration map, and validation failure threshold without retraining or recalibration.

We evaluate restoration using RMSE, MAE, Pearson correlation, SNR improvement, R-peak F1, and heart-rate MAE. R-peaks are extracted from clean and restored ECG windows using a standard QRS detection procedure~\cite{4122029}. Reliability is evaluated using Spearman correlation with true restoration risk, AUROC/AUPRC for failures defined by the validation 80th-percentile true risk, risk expected calibration error, and selective ECG analysis at 100\%, 95\%, 90\%, 80\%, and 70\% coverage. For PTB-XL, we report ranking and selective-analysis metrics with no PTB-XL-specific calibration.

\begin{figure*}[!t]
    \centering
    \begin{minipage}{0.44\textwidth}
        \centering
        \includegraphics[width=\linewidth]{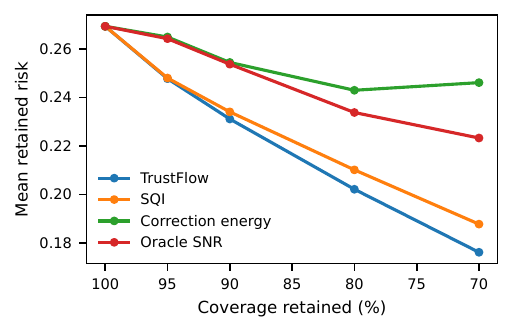}
        \vspace{-0.6em}
        \centerline{\footnotesize (a) Coverage-risk curve}
    \end{minipage}
    \hfill
    \begin{minipage}{0.44\textwidth}
        \centering
        \includegraphics[width=\linewidth]{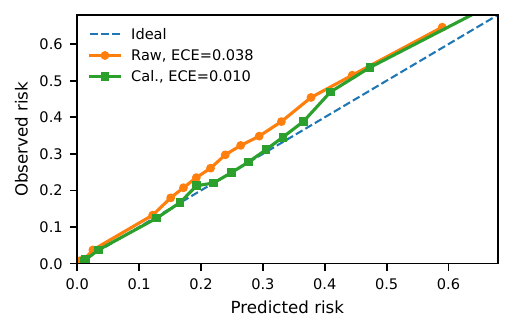}
        \vspace{-0.6em}
        \centerline{\footnotesize (b) Risk calibration}
    \end{minipage}
        \caption{Reliability behavior of TrustFlow-ECG. Lower predicted-risk windows have lower observed restoration risk, and validation-set calibration improves alignment between predicted and observed risk.}
    \label{fig:coverage_calibration}
    \vspace{-0.6em}
\end{figure*}

\section{Results and Discussion}
\label{sec:results}

\subsection{Restoration performance}
\label{subsec:restoration_results}

Table~\ref{tab:restoration_summary} reports restoration performance on the fixed test set. The conditional flow model gives the best waveform and beat-level results, so we use it as the restoration backbone for the reliability experiments.

\begin{table}[t]
\centering
\caption{Restoration performance on the fixed test set. Lower is better for RMSE, MAE, and HR MAE; higher is better otherwise.}
\label{tab:restoration_summary}
\scriptsize
\setlength{\tabcolsep}{3.0pt}
\begin{tabular}{lcccccc}
\hline
\textbf{Model} & \textbf{RMSE} $\downarrow$ & \textbf{MAE} $\downarrow$ & \textbf{Pearson} $\uparrow$ & \textbf{F1} $\uparrow$ & \textbf{HR MAE} $\downarrow$ & \textbf{$\Delta$SNR} $\uparrow$ \\
\hline
U-Net & 0.2952 & 0.2234 & 0.9415 & 0.8254 & 7.85 & 3.17 \\
Transformer & 0.2846 & 0.2206 & 0.9412 & 0.8274 & 9.19 & 2.48 \\
\textbf{Flow} & \textbf{0.1848} & \textbf{0.1468} & \textbf{0.9707} & \textbf{0.8778} & \textbf{6.26} & \textbf{10.40} \\
\hline
\end{tabular}
\end{table}

\subsection{Reliability prediction and calibration}
\label{subsec:reliability_results}

Table~\ref{tab:reliability_summary} compares scores for predicting true restoration risk. Correction energy gives weak ranking, oracle SNR is unavailable at inference, and the handcrafted SQI regressor is competitive. TrustFlow gives the best Spearman correlation, AUROC, and AUPRC, while calibration reduces ECE from 0.0385 to 0.0098 with little loss in ranking.

\begin{table}[t]
\centering
\caption{Reliability prediction results. Failures use the validation 80th-percentile true risk. Oracle SNR is not available at inference.}
\label{tab:reliability_summary}
\scriptsize
\setlength{\tabcolsep}{4.0pt}
\begin{tabular}{lcccc}
\hline
\textbf{Score} &
\textbf{Spearman} $\uparrow$ &
\textbf{AUROC} $\uparrow$ &
\textbf{AUPRC} $\uparrow$ &
\textbf{ECE} $\downarrow$ \\
\hline
Correction energy & 0.2454 & 0.7048 & 0.4342 & 0.0488 \\
Oracle SNR & 0.4565 & 0.8003 & 0.5418 & -- \\
Handcrafted SQI & 0.8157 & 0.8703 & 0.7324 & 0.0119 \\
TrustFlow risk & \textbf{0.8615} & \textbf{0.9105} & \textbf{0.8188} & 0.0385 \\
TrustFlow risk + cal. & 0.8608 & 0.9092 & 0.8140 & \textbf{0.0098} \\
\hline
\end{tabular}
\end{table}

\begin{table}[t]
\centering
\caption{Reliability-input ablation. The full input combines the noisy ECG, restored ECG, and correction residual.}
\label{tab:input_ablation}
\scriptsize
\setlength{\tabcolsep}{4.0pt}
\begin{tabular}{lcccc}
\hline
\textbf{Input} &
\textbf{Spearman} $\uparrow$ &
\textbf{AUROC} $\uparrow$ &
\textbf{AUPRC} $\uparrow$ &
\textbf{ECE} $\downarrow$ \\
\hline
Noisy only & 0.7508 & 0.8473 & 0.6403 & 0.0242 \\
Restored only & 0.3201 & 0.6778 & 0.5051 & 0.0463 \\
Noisy + restored & 0.8084 & 0.8695 & 0.7260 & 0.0370 \\
\textbf{Noisy + restored + residual} &
\textbf{0.8608} & \textbf{0.9092} & \textbf{0.8140} & \textbf{0.0098} \\
\hline
\end{tabular}
\end{table}

Table~\ref{tab:input_ablation} confirms the value of the correction residual; Fig.~\ref{fig:coverage_calibration} shows lower observed risk for lower predicted risk and improved alignment after calibration.

\subsection{Selective ECG analysis}
\label{subsec:selective_results}

Table~\ref{tab:selective_analysis} reports selective ECG analysis using calibrated risk. At 80\% coverage, RMSE decreases from 0.1848 to 0.1345, R-peak F1 increases from 0.8778 to 0.9175, and HR MAE decreases from 6.26 to 4.88 bpm. Bootstrap results over 1{,}000 window-level replicates were stable: RMSE was 0.1345 [0.1303, 0.1391] and R-peak F1 was 0.9175 [0.9133, 0.9213].

\begin{table}[t]
\centering
\caption{Selective ECG analysis using calibrated TrustFlow risk. HR denotes HR MAE in bpm.}
\label{tab:selective_analysis}
\scriptsize
\setlength{\tabcolsep}{4.0pt}
\begin{tabular}{ccccccc}
\hline
\textbf{Cov.} &
\textbf{N} &
\textbf{RMSE} $\downarrow$ &
\textbf{MAE} $\downarrow$ &
\textbf{Pearson} $\uparrow$ &
\textbf{F1} $\uparrow$ &
\textbf{HR} $\downarrow$ \\
\hline
100\% & 4000 & 0.1848 & 0.1468 & 0.9707 & 0.8778 & 6.26 \\
95\%  & 3800 & 0.1662 & 0.1325 & 0.9771 & 0.8925 & 5.82 \\
90\%  & 3600 & 0.1530 & 0.1218 & 0.9815 & 0.9040 & 5.40 \\
80\%  & 3200 & 0.1345 & 0.1064 & 0.9862 & 0.9175 & 4.88 \\
70\%  & 2800 & 0.1172 & 0.0920 & 0.9895 & 0.9290 & 4.35 \\
\hline
\end{tabular}
\end{table}

Selective gains were not driven only by easy artifact categories. Without powerline examples, RMSE decreased from 0.2293 to 0.1835, R-peak F1 improved from 0.8519 to 0.8915, and HR MAE decreased from 7.62 to 6.35 bpm. On the EMG+motion+mixed subset, RMSE decreased from 0.2351 to 0.1885, R-peak F1 improved from 0.8454 to 0.8870, and HR MAE decreased from 7.63 to 6.40 bpm.

On the external PTB-XL stress set, the model retained strong ranking without retraining, with Spearman 0.753 and AUROC 0.866. At 80\% coverage, selective rejection reduced RMSE from 0.2857 to 0.2441, increased R-peak F1 from 0.9216 to 0.9340, and reduced HR MAE from 10.48 to 9.34 bpm.

\section{Conclusion}
\label{sec:conclusion}

We presented TrustFlow-ECG, a reliability-aware ECG restoration framework that estimates calibrated window-level risk from inference-time signals. Results on corrupted MIT-BIH/NSTDB windows and an external PTB-XL stress check show improved risk ranking, calibration, and selective retained-window quality without clean ECG at inference. Future work will evaluate naturally noisy ambulatory ECG where paired clean references are unavailable.

\section*{Compliance with Ethical Standards}
This study used publicly available, de-identified ECG datasets with simulated artifact corruption. No ethical approval was required. No funding was received. The authors have no relevant financial or nonfinancial interests to disclose.

\bibliographystyle{IEEEbib}
\bibliography{refs}

\end{document}